\documentclass[conference]{IEEEtran}
\usepackage{cite}

\ifCLASSINFOpdf
  \usepackage[pdftex]{graphicx}
\else
\fi
\usepackage{amsmath}
\usepackage{algorithmic}

\usepackage{array}
\usepackage{url}

\usepackage{xcolor}
\begin{document}
%
\title{Mobility-Aware Smart Charging of Electric \\Bus Fleets}



%
\author{\IEEEauthorblockN{Ahmadreza Moradipari\IEEEauthorrefmark{2}\IEEEauthorrefmark{1},
Nathaniel Tucker\IEEEauthorrefmark{2}\IEEEauthorrefmark{1},
Tuo Zhang\IEEEauthorrefmark{2}, 
Gustavo Cezar\IEEEauthorrefmark{3} and
Mahnoosh Alizadeh\IEEEauthorrefmark{2}}
\IEEEauthorblockA{\IEEEauthorrefmark{2}Department of Electrical and Computer Engineering, 
University of California, Santa Barbara,
California, 93106, USA}
\IEEEauthorblockA{\IEEEauthorrefmark{3}SLAC National Accelerator Laboratory, GISMo Group, California, 94025, USA}
\IEEEauthorblockA{\IEEEauthorrefmark{1}Authors have equal contribution}
}


\maketitle

\begin{abstract}
We study the joint route assignment and charge scheduling problem of a transit system dispatcher operating a fleet of electric buses in order to maximize solar energy integration and reduce energy costs. Specifically, we consider a complex bus transit system with preexisting routes, limited charging infrastructure, limited number of electric buses, and time-varying electricity rates. We present a mixed integer linear program (MILP) that yields the minimal cost daily operation strategy for the fleet (i.e., route assignments and charging schedules using daily solar forecasts). We present numerical results from a real-world case study with Stanford University's Marguerite Shuttle (a large-scale electric bus fleet) to demonstrate the validity of our solution and highlight the significant cost savings compared to the status quo.
\end{abstract}


%
\IEEEpeerreviewmaketitle
\makeatletter
\def\blfootnote{\xdef\@thefnmark{}\@footnotetext}
\makeatother

\section{Introduction}
\label{section: introduction}

Due to the potential reduction in operational costs \cite{Intro_cost_reduction}, elimination of tailpipe emissions \cite{Intro_pollute_reduction}, and encouragement from government agencies \cite{Intro_government}, transit systems have started to purchase electric buses over the traditional diesel or compressed natural gas (CNG) buses. At surface level, replacing traditional buses with electric buses might seem like a simple task; however, there are many obstacles preventing a transit system from simply assigning electric buses to existing routes that were previously served by diesel buses. 

The two most fundamental obstacles are the restricted travel distance and lengthy recharge time of electric buses. Even with recent advances in electric transportation and battery technology, modern electric buses are commonly restricted to operate within 20\%-95\% state of charge (SOC) to prevent stressing the batteries and reducing lifespan \cite{Intro_battery_range}. Combining this SOC limitation with the high cost of large battery packs, most electric buses are currently inferior to diesel/CNG buses in operational range. Second, the recharging process of an electric bus takes significantly more time than the refueling process of a diesel/CNG bus\cite{Intro_battery_range}. Additionally, due to the lengthy recharge time and limited charging infrastructure, the transit system dispatcher must be mindful of how the fleet's recharging infrastructure is managed in order to provide adequate energy to serve routes. 

Despite the aforementioned challenges, the promise of eliminating large amounts of greenhouse gas emissions from transit buses has enticed early adopters to operate fleets of electric buses since the early 21st century \cite{Intro_cost_reduction}; however, it is likely that these electric bus fleets are operating suboptimally in their recharging strategies and route assignments \cite{ROGGE2018}. Accordingly, there has been increasing   interest in the optimal operation and infrastructure planning of electric bus fleets.

The first category of work that studies optimized charging for electric bus fleets considers the assignment of buses to routes as given, i.e., the times at which each bus is parked and is available to recharge is predetermined. Specifically, the authors of \cite{kunith_placing_sizing_02} present an optimization model for installing charging infrastructure and sizing batteries for a cost-effective electric bus fleet. Similarly, the authors of \cite{ROGGE2018} consider infrastructure planning as well as fleet composition and the recharging process, with the goal of minimizing total cost of ownership (TOC) of the fleet. Moving away from infrastructure planning, the authors of \cite{houbbadi_overnight_04} present a method to minimize battery aging costs of an electric bus fleet recharging at nighttime. The authors of \cite{qin2016numerical} present the cost savings from controlling the charging thresholds for a fleet of electric buses serving one route continuously in Tallahassee, Florida. Similarly, the authors of \cite{wang2017optimal} present a MILP framework for scheduling bus charging and show the potential cost savings from an electric bus fleet in Davis, California. Furthermore, \cite{chen_coordingated_05} presents a charging strategy for electric buses with fast charging infrastructure.

Considering both route assignment and charge scheduling (i.e., the mobility-aware setting) the authors of \cite{paul_yamada_06} present a k-greedy solution method to maximize travel distance of each electric bus within the fleet. A work similar to ours, \cite{exact_07}, presents a linear formulation for route assignment and charge scheduling; however, the aim is to minimize the number of electric buses needed to replace an existing diesel fleet. Hence, the variability of electricity costs are not considered. 

Similar to the aforementioned papers, the work presented in this manuscript considers both the route assignment and charge scheduling problem of an electric bus fleet. However, the presented approach is able to improve upon previous mobility-aware work by accounting for time-varying electricity prices, utilizing on-site solar energy generation, and providing a minimal cost schedule for the fleet's daily operation.

\textit{Organization:} Section \ref{section: problem description} describes the problem of a fleet dispatcher operating a fleet of electric buses and proposes a mixed integer linear program (MILP) formulation that solves for the minimal cost route assignments and recharging schedule. Section \ref{section: case study} presents the results of the MILP for the real-world example of Stanford's Marguerite Shuttle Transit System.

\begin{figure}
\label{fig: map}
\centering
\includegraphics[width=\columnwidth]{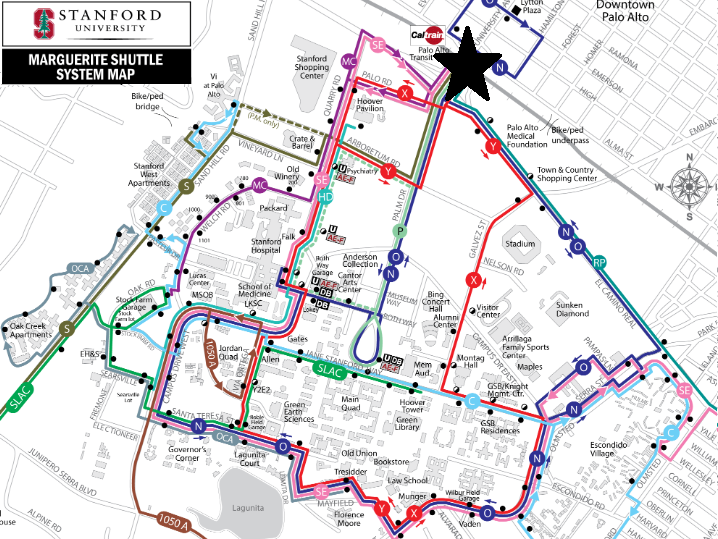}
\caption{Primary service area for Stanford University's Marguerite Shuttle. Trip origins at Caltrain Palo Alto Transit Center (star). Full system map available at: https://transportation.stanford.edu/marguerite}
\label{fig_sim}
\end{figure}

\section{Problem Description}
\label{section: problem description}

We consider a fleet dispatcher attempting to optimize an electric bus transit system. Specifically, the fleet dispatcher aims to assign electric buses to serve the daily trips and schedule the recharging of the buses to minimize electricity cost (e.g., recharging during the inexpensive electricity rates of nighttime or when solar generation is abundant while still fulfilling all required bus routes). In the following, we consider the case where the physical infrastructure (e.g., buses, chargers, parking spots, etc.) and time-tables (e.g., routes, stops, start/end times, etc.) are already established within the transit system, but not yet optimized for the aforementioned objective (as is the case for the Stanford University Marguerite Shuttle, discussed in Section \ref{section: case study}). Given the transit system's fixed time-table and electric bus infrastructure, the fleet dispatcher seeks to answer questions such as the following:
\begin{enumerate}
    \item Which electric bus should be assigned to each route at each time?
    \item When should each electric bus be recharged?
    \item Does the system need to utilize spare diesel buses to supplement the electric buses? 
    \item Would more infrastructure benefit the daily operation of the electric bus fleet?
    \item What size of on-site solar generation system is needed to fully supply the fleet with renewable energy?
\end{enumerate}

Let us consider the Stanford Marguerite Shuttle Transit System (Figure \ref{fig: map}) which consists of 38 electric buses, 23 diesel buses, 23 electric bus chargers, and total of 20 daily routes.
Currently, the assignment of buses to routes and their recharging strategy follows rules adopted by operators that work well in practice by ensuring sufficient charge is available for service. However, as we demonstrate in our numerical case study, the current assignment results in significant losses for the transit system in terms of daily operational costs and can be improved upon through a joint charge and route assignment policy. As such, in order to optimize the decision making problem of the fleet dispatcher, we formulate a  mixed-integer-linear-program (MILP) to solve for both the optimal recharging schedules and route assignments for an electric bus transit system.

\subsection{MILP Formulation}
\label{section: MILP formulation}
In the electric bus transit system, we consider one central transit center (i.e., bus depot) from which all the buses start and finish their routes as well as recharge. The buses are required to serve numerous routes throughout the service area, and each route must be served multiple times each day (i.e., the electric bus fleet is required to fulfill multiple \textit{trips} for each route). We denote $\mathcal{S}$ as the set of scheduled trips across all routes that need to be fulfilled. For each trip $i\in\mathcal{S}$, let $a_i$ and $b_i$ denote the start and end time of trip $i$. More specifically, these are the times that a bus leaves the depot and later returns if serving trip $i$. If trip $i$ is a one-way route that does not loop back to the depot, we account for the extra duration for the bus to return to the depot in $b_i$ accordingly (i.e., the trip end time $b_i$ accounts for ``deadhead" travel). Similarly, if a route does not start at the depot, we account for the deadhead travel time to the starting location in $a_i$. 

In order to capture the state of charge of each bus at any time $t$, we discretize the day into $T$ time steps (e.g., five minute intervals) and $\mathcal{T}$ is the set of time steps for an entire day. Furthermore, let $d_i$ be the  energy consumption per time step for a bus serving trip $i$  (while we assume that varying traffic conditions across different routes can affect energy consumption rates, we assume that the buses are identical in their energy consumption when they serve the same route). Let $\mathcal{K}$ be the set of electric buses and $\mathcal{N}$ be the set of electric bus chargers installed at the central depot. For each charger $n \in \mathcal{N}$, $u_n$ is the charging rate. Additionally, let ${\bf p} = [p(t)]_{t \in \mathcal{T}}$ be the vector of electricity prices for an  entire day. We denote as $E^k_{min}$ and $E^k_{max}$ the minimum and maximum energy levels for bus $k$, respectively. The fleet dispatcher usually sets $E^k_{min} > 0, \forall k\in\mathcal{K}$ for safety precautions. {Let $g(t)$ be the available on-site solar generation at time $t$, which we assume is known at the time of dispatch.} Moreover, we assume that the electricity used from the on-site solar generation  is free for the operator. Last, we denote the initial energy level of bus $k$ as $e_0^k$. 

Next, we describe the decision variables used in the MILP formulation. We set the binary variable $X_i^k(t)$ to $1$ if bus $k$ is serving trip $i$ at time $t$ and $0$ otherwise. We set the binary variable $Z_k(t)$ to $1$ if bus $k$ is charging at time $t$ and $0$ otherwise. We set the binary variable $Y_n^k(t)$ to $1$ if bus $k$ is occupying charger $n$ at time $t$ and $0$ otherwise. We use the variable $E^k(t)$ to track the energy level of bus $k$ at time $t$. {Lastly, let $V(t)$ be the total amount of electricity that the dispatcher purchases from the grid at time $t$, and $S(t)$ be the amount of electricity that buses obtain from the available on-site solar generation at time $t$.} With the necessary notation and decision variables, the joint charging and routing MILP for the electric bus fleet can be formulated as follows:

\begin{subequations}
\begin{flalign}
    \label{eqn: objective}
    &\text{Minimize } \hspace{10pt} \sum_{t\in\mathcal{T}} \hspace{10pt} p(t) V(t)&&\\
    &\nonumber\text{Subject to:}&&
\end{flalign}
\vspace{-4.5ex}
\begin{flalign}
    \label{eqn: charge_or_trip constraint}
    &Z^k(t) + \sum_{i\in\mathcal{S}}X^k_i(t) \leq 1, &&\hspace{-5pt}\forall k\in\mathcal{K}, t\in\mathcal{T}\\
    \label{eqn: trip_has_to_be_served constraint}
    &\sum_{k\in\mathcal{K}} X_i^k(t) = 1, &&\hspace{-5pt}\forall i\in\mathcal{S}, t\in[a_i,b_i]\\
    \label{eqn: cant_switch_buses_on_trip constraint}
    &X_i^k(t+1) = X_i^k(t), &&\hspace{-5pt}\forall i\in\mathcal{S}, k\in\mathcal{K}, t\in[a_i,b_i\hspace{-2pt}-\hspace{-2pt}1]\\
    \label{eqn: use_one_charger constraint}
    &\sum_{k\in\mathcal{K}} Y_n^k(t) \leq 1, &&\hspace{-5pt}\forall n\in\mathcal{N},t\in\mathcal{T}\\
    \label{eqn: charging_at_charger constraint}
    &\sum_{n\in\mathcal{N}}Y_n^k(t)=Z^k(t), &&\hspace{-5pt}\forall k\in\mathcal{K},t\in\mathcal{T}
\end{flalign}
\vspace{-3ex}
\begin{flalign}
    \label{eqn: energy_level constraint}
    E^k(t) &= E^k(t-1) + \sum_{n\in\mathcal{N}} u_n Y_n^k(t) - \sum_{i\in\mathcal{S}} d_i X_i^k(t),&&\\
    \nonumber&  \hspace{89pt} \forall k\in\mathcal{K}, t\in\mathcal{T}
\end{flalign}
\vspace{-3.5ex}
\begin{flalign}
    \label{eqn: charging from grid or solar}
    &\sum_{n\in\mathcal{N}} \sum_{k\in\mathcal{K}} Y_n^k(t) u_n =V(t) + S(t), &&\hspace{-2.5pt}\forall t\in\mathcal{T} \\
    \label{eqn: energy_threshold constraint}
    &E^k_{min} \leq E^k(t) \leq E^k_{max}, &&\hspace{-2pt}\forall k\in\mathcal{K}, t\in\mathcal{T}\hspace{58pt}\\
    \label{eqn: x_binary constraint}
    &X_i^k(t) \in \{0,1\}, &&\hspace{-2pt}\forall i\in\mathcal{S}, k\in\mathcal{K}, t\in\mathcal{T}\\
    \label{eqn: y_binary constraint}
    &Y_n^k(t) \in \{0,1\}, &&\hspace{-2pt}\forall n\in\mathcal{N}, k\in\mathcal{K}, t\in\mathcal{T}\\
    \label{eqn: z_binary constraint}
    &Z^k(t) \in \{0,1\}, &&\hspace{-2pt}\forall k\in\mathcal{K}, t\in\mathcal{T}\\
    \label{eqn: solar.constraint}
    & 0 \leq S(t) \leq g(t), &&\hspace{-2pt} \forall t\in\mathcal{T}\\
    \label{eqn: initial_energy constraint}
    &E^k(0) = e_0^k, &&\hspace{-2pt}\forall k\in\mathcal{K}\\
    \label{eqn: final_energy constraint}
    &E^k(T) = e_0^k, &&\hspace{-2pt}\forall k\in\mathcal{K}.
\end{flalign}
\end{subequations}

The objective in equation \eqref{eqn: objective} aims to minimize the daily electricity cost of recharging the bus fleet. Constraint \eqref{eqn: charge_or_trip constraint} ensures that a bus is either charging, serving a trip, or parked in the depot (without charging). Constraint \eqref{eqn: trip_has_to_be_served constraint} ensures that all the required daily trips will be served by a bus. Constraint \eqref{eqn: cant_switch_buses_on_trip constraint} ensures that one unique bus will serve each trip (i.e., a trip cannot be interrupted to switch buses). Constraint \eqref{eqn: use_one_charger constraint} ensures that a bus can only occupy one charger per time slot. Constraint \eqref{eqn: charging_at_charger constraint} guarantees that if a bus is occupying a charger, then it is charging. Constraint \eqref{eqn: energy_level constraint} calculates the energy level of each bus in each time epoch. Specifically, the energy level at time $t$ is equal to the energy level at time $t-1$ plus the charged energy if the bus was charging or minus the spent energy if the bus was serving a trip. { Constraint \eqref{eqn: charging from grid or solar} ensures that buses obtain electricity from either the grid or on-site solar.}
Constraint \eqref{eqn: energy_threshold constraint} ensures that the buses operate above a desired minimum energy threshold. Constraints \eqref{eqn: x_binary constraint}-\eqref{eqn: z_binary constraint} are binary constraints on the decision variables. { Constraint \eqref{eqn: solar.constraint} ensures that the solar energy used by the bus fleet is less than or equal to available solar generation at time $t$. } Lastly, constraint \eqref{eqn: initial_energy constraint} sets the initial energy of each bus and constraint \eqref{eqn: final_energy constraint} ensures that the energy level of the fleet returns to the initial value so the same route assignments and charge schedule can be used for the next day.

\subsection{Behind-the-Meter Solar Integration}
\label{section: solar}

To exploit free on-site solar energy and to avoid injecting excess power back into the distribution grid, the fleet dispatcher prioritizes recharging the buses during periods when solar generation is available. Only if there is not enough solar energy, then the fleet dispatcher should purchase electricity from the grid. As stated in Section \ref{section: MILP formulation}, to accommodate behind-the-meter solar integration, the dispatcher's MILP formulation makes use of a daily solar forecast, $g(t)|_{t=1,\dots,T}$. This can be estimated from forecast models, including those that use weather forecasts, and previous years' solar irradiance data. We note that if the solar generation is over-estimated, then the fleet will have to purchase more expensive grid energy potentially during peak times such as midday. As such, a conservative estimate is preferred as cheaper electricity can be procured in the late night period. Future work could investigate moving-horizon solution methods to account for stochastic solar generation and update the route and charge assignments in real-time as solar energy data becomes available.

\section{Case Study}
\label{section: case study}

As stated in the introduction, the motivation for the proposed MILP for electric bus fleets is the real-world Stanford Marguerite Shuttle Transit System (Figure \ref{fig: map}). The Marguerite Shuttle System is free, open to the public, and operates seven days a week all year traversing the Stanford campus and surrounding areas. More specific information can be found at https://transportation.stanford.edu/marguerite.

\subsection{Stanford Marguerite Shuttle System Information}
\label{section: marguerite information}

\begin{figure}
\centering
\includegraphics[width=\columnwidth]{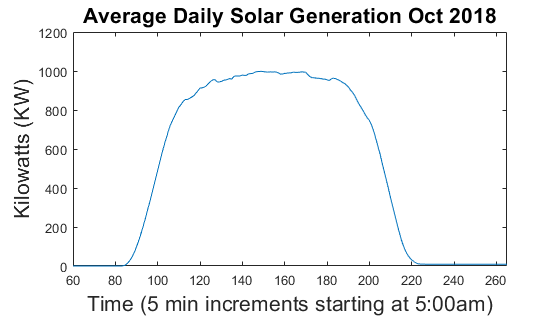}
\caption{Average daily solar generation for a 1 MW on-site installation. Data averaged from CAISO renewable database in October 2019.}
\label{fig: solar}
\end{figure}

Currently, the Marguerite fleet consists of 23 diesel buses and 38 electric buses from BYD split into 10 K7 models with battery capacity of 197kWh, 10 K9 models and 18 K9M models, both with 324kWh battery capacity. Additionally, the central depot is equipped with 23 double port electric bus chargers where each port can deliver up to 40kW. Each bus can be charged from one or two ports for a total power of 80kW. For the electricity rates, we consider PG\&E's E-20 electricity rate structure for off-peak, partial-peak, and peak hours. The electricity rates are given in Table \ref{table: electricity_price}.
\begin{table}
\renewcommand{\arraystretch}{1.2}
\caption{PG\&E E-20 Rate Structure}
\label{table: electricity_price}
\centering
\begin{tabular}{|l|l|l|}
\hline
\textbf{Time Interval} & \textbf{Label} &\textbf{Price}\\
\hline
12:00am-8:30am & Off-Peak & \$0.08422/kWh \\
\hline
8:30am-12:00pm & Partial-Peak & \$0.11356/kWh\\
\hline
12:00pm-6:00pm & Peak & \$0.16127/kWh\\
\hline
6:00pm-9:30pm & Partial-Peak & \$0.11356/kWh\\
\hline
9:30pm-12:00am & Off-Peak & \$0.08422/kWh \\
\hline
\end{tabular}
\end{table}
Furthermore, the Marguerite Shuttle system serves up to 20 unique routes on any given day. Across all 20 routes, 15 of them are mainly fulfilled by electric buses, meaning that the electric bus fleet is required to make 352 trips per day, during weekdays. The specific routes and mileages are listed in Table \ref{table: route information}. For the purposes of this numerical example, the solar forecast used was an average daily solar generation calculated from October 2019 with a maximum generation of 1 MW. The solar forecast is displayed in Figure \ref{fig: solar}. 
\begin{table}
\renewcommand{\arraystretch}{1.2}
\caption{Stanford Marguerite Shuttle Route Information}
\label{table: route information}
\centering
\begin{tabular}{|l|c|c|}
\hline
\textbf{Route Name} & \textbf{Daily Trips} &\textbf{Trip Miles}\\
\hline
C Line & 33 & 7.00 \\
\hline
C Limited & 11 & 4.60 \\
\hline
MC Line (AM/PM)& 46 & 3.00 \\
\hline
MC Line (Mid Day) & 11 & 5.10 \\
\hline
P Line (AM/PM)	&56	&2.50\\
\hline
P Line (Mid Day)	&11	&4.00\\
\hline
Research Park (AM/PM)	&24	&10.40\\
\hline
X Express (AM)	&12	&1.20\\
\hline
X Line	&44	&4.60\\
\hline
X Limited (AM)	&10	&2.00\\
\hline
X Limited (PM)	&10	&1.50\\
\hline
Y Express (PM)	&20	&1.20\\
\hline
Y Line	&44	&4.60\\
\hline
Y Limited (AM)	&10	&2.40\\
\hline
Y Limited (PM)	&10	&2.00\\
\hline
\textbf{Totals} & \textbf{352 trips/day} & \textbf{1431.50 miles/day}\\
\hline
\end{tabular}
\end{table}

\subsection{Simulation Results}
\label{section: simulation results}
The proposed MILP was implemented in Matlab making use of CVX and Mosek. All numerical experiments were run on a laptop with 16 GB of RAM and 3.5 GHz Intel i7 processor. This section reports on the charging schedule, route assignments, and cost savings when comparing the proposed MILP solution with on-site solar generation, without on-site solar generation, and the status quo (i.e., the status quo is the actual operations of the Stanford Marguerite Fleet from 7-October-2019) which does not yet exploit free on-site solar generation.

\begin{figure}
\centering
\includegraphics[width=\columnwidth]{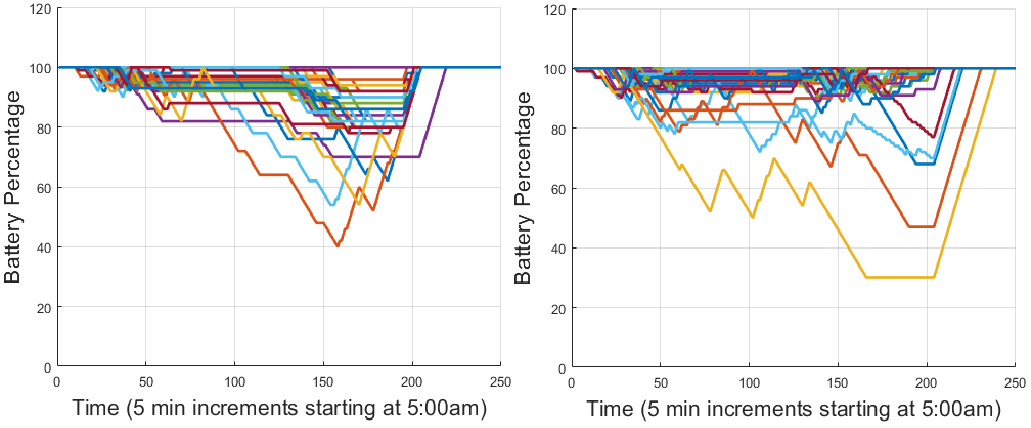}
\caption{Left: Battery levels for each electric bus when considering a fleet without available on-site solar. Right: Battery levels for each electric bus when optimizing with available on-site solar generation.}
\label{fig: energy_each_bus}
\end{figure}

\begin{figure}
\centering
\includegraphics[width=\columnwidth]{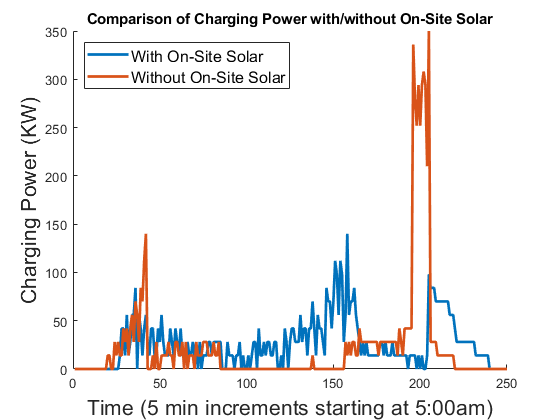}
\caption{Total charging power of the fleet throughout the day. Blue: Solution accounting for on-site solar generation. Red: Solution does not include on-site solar generation.}
\label{fig: total_charging_power}
\end{figure}

Figure \ref{fig: energy_each_bus} presents the energy levels of each bus in the fleet during the day when the dispatch is generated through our proposed MILP. Time on the x-axis begins at 5:00am, as this is the start of the earliest route that must be fulfilled. The left plot shows the energy levels of the buses when the MILP is not utilizing on-site solar generation. The right plot shows the battery levels of the buses when the MILP accounts for on-site solar generation. It will become more clear when examining Figure \ref{fig: total_charging_power} that the buses charge more during midday in the right plot than the left, to make use of the free on-site solar.

Figure \ref{fig: total_charging_power} presents the total charging power of the fleet across the entire day. The red curve presents the total charging power for the MILP solution that does not exploit on-site solar generation. Conversely, the blue plot shows the fleet's total charging power from the MILP solution that does account for on-site solar generation. It is clear from this plot that the solution that accounts for on-site solar (blue) is able to charge in the middle of the day when solar is abundant; however, the solution that does not exploit solar (red) does not charge during the midday as the electricity prices are highest at this time. Instead, the fleet has a spike in charging power in the evening when electricity rates are decreased. This large transient in the evening could be detrimental to grid stability, increase in harmonics, accelerate aging of grid assets (i.e. transformers) and could potentially lead to demand charges for the fleet dispatcher due to high power consumption. As such, the solution making use of on-site solar generation with a forecasting method is preferable.

\begin{figure}
\centering
\includegraphics[width=\columnwidth]{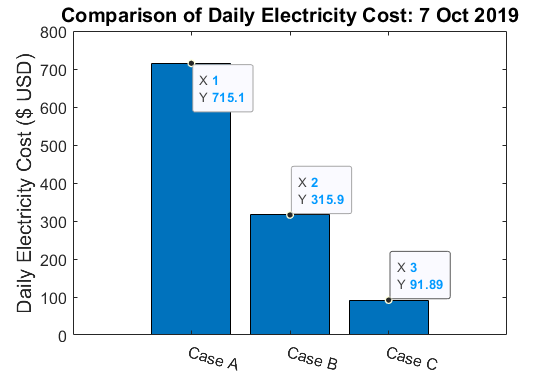}
\caption{Price Comparison for 3 difference regimes: Case 1: Status Quo, electric bus charging data obtained from real-implementation (Stanford Marguerite Shuttle) on 7-Oct-2019. Case 2: Mobility-Aware MILP solution for same routes and buses as Case A, \textit{without} on-site solar generation. Case 3: Mobility-Aware MILP solution for same routes and buses as Case A, \textit{with} on-site solar generation.}
\label{fig: price_compare}
\end{figure}

Last, Figure \ref{fig: price_compare} presents the daily electricity costs for the three different test cases. Case A: Status Quo. We had access to the data from the operations of the Stanford Marguerite fleet on 7-October-2019 and calculated the cost of charging the fleet under the E-20 rate structure. As such, under normal operation, the daily operational cost was \$715.10 USD. Case B corresponds to the solution of the proposed MILP with the same routes, buses, and chargers as Case A; however, the mobility-aware solution reassigned buses to new trips and rescheduled the charging of each bus. In Case B, the MILP solution did not account for on-site solar and the daily cost was \$267.90 USD. Last, Case C was identical to Case B; however, the MILP accounted for the on-site solar generation and had access to the daily solar forecast. As such, the daily cost was reduced to \$61.89 USD. From these results, it is evident that the fleet dispatcher benefits from the MILP formulation for routing and charging ($55\%$ decrease in cost in Case B).

\section{Conclusion}
In this paper, we investigated the joint route assignment and charge scheduling problem of a transit system dispatcher operating a fleet of electric buses in order to maximize solar energy integration and reduce energy costs. We considered a complex bus transit system with preexisting routes, limited charging infrastructure, limited number of electric buses, and time-varying electricity rates. We presented a mixed integer linear program (MILP) that yields route assignments and charging schedules using daily solar forecasts. We presented numerical results from a real-world case study with Stanford University's Marguerite Shuttle to demonstrate the cost-saving benefits of our solution and highlight the significant cost savings compared to the status quo. 

Future work includes investigating a moving-horizon solution approach to account for stochastic solar generation. Additionally, we would like to add traditional diesel routes to the optimization to further minimize emissions and to expand the clean operation of the electric bus fleet. Further future work can include performing field test experiments with real buses during operational hours, determining the optimal solar capacity to fully charge the electric bus fleet, and quantify the value and size of onsite solar and battery combination for resiliency.

\section*{Acknowledgment}
The authors would like to thank the Stanford Transportation team for the support, discussions, and information about operations.
This work was funded by the California Energy Commission
under grant EPC-17-020. SLAC National Accelerator Laboratory is operated for the US Department of Energy by Stanford
University under Contract DE-AC02-76SF00515.



%

\bibliographystyle{IEEEtran}
\bibliography{references}


\end{document}